# National Access Points for Intelligent Transport Systems Data: From Conceptualization to Benefits Recognition and Exploitation

Georgia Aifantopoulou[1], Chrysostomos Mylonas[2], Alexandros Dolianitis[3], Afroditi Stamelou[4], Vasileios Psonis[5], Evangelos Mitsakis[6]

[1]PhD, Deputy Director-Research Director, Head of Sector B, Scientist A, Centre for Research and Technology Hellas, Hellenic Institute of Transport, Thermi, Thessaloniki
E-mail: gea@certh.gr

[2]Research Associate, Centre for Research and Technology Hellas, Hellenic Institute of Transport, Thermi, Thessaloniki
E-mail: chmylonas@certh.gr

[3]Research Associate, Centre for Research and Technology Hellas, Hellenic Institute of Transport, Thermi, Thessaloniki
E-mail: adolianitis@certh.gr

[4]Research Associate, Centre for Research and Technology Hellas, Hellenic Institute of Transport, Thermi, Thessaloniki
E-mail: stamelou@certh.gr

[5]PhD, Research Associate, Centre for Research and Technology Hellas, Hellenic Institute of Transport, Thermi, Thessaloniki
E-mail: psonis@certh.gr

[6]PhD, Senior Researcher, Scientist B, Centre for Research and Technology Hellas, Hellenic Institute of Transport, Thermi, Thessaloniki
E-mail: emit@certh.gr

## Abstract

Intelligent Transport Systems (ITS) constitute a core representative of a paradigm shift in the transport sector. The extent to which the transport sector has adapted itself to this digital era relies considerably on the availability of suitable and reliable data. Currently, several data-related limitations, such as the scarcity of available datasets, hinder the deployment of ITS services. Such limitations may be overcome with the deployment of properly designed data exchange platforms that enable a seamless life-cycle of data harvesting, processing, and sharing. The European Union recognizing the potential benefits of such platforms has, through the relevant Delegated Regulations, proposed the development of a National Access Point (NAP) by each individual Member State. This paper aims to ascertain the role of a NAP within the ITS ecosystem, to investigate methodologies used in designing such platforms, and, through the drafting of an extended use case, showcase a NAP's operational process and associate possible benefits with specific steps of it.

## Περίληψη

Τα Ευφυή Συστήματα Μεταφορών (ΕΣΜ) αποτελούν ένα από τους βασικούς αντιπρόσωπους των εκτεταμένων αλλαγών που παρατηρούνται στον τομέα των Μεταφορών. Ο βαθμός στον οποίο ο τομέας των Μεταφορών έχει προσαρμοστεί στις απαιτήσεις της ψηφιακής εποχής, βασίζεται σημαντικά στην διαθεσιμότητα κατάλληλων και αξιόπιστων δεδομένων. Επί του παρόντος, παρατηρούνται αρκετοί περιορισμοί ως προς τα δεδομένα αυτά, οι οποίοι υποδαυλίζουν την περαιτέρω ανάπτυξη και παροχή υπηρεσιών ΕΣΜ. Οι εν λόγω περιορισμοί μπορούν να υπερσκελιστούν μέσω της ανάπτυξης και λειτουργίας ορθώς σχεδιασμένων πλατφορμών ανταλλαγής δεδομένων. Τέτοιες πλατφόρμες εξασφαλίζουν έναν συνεχή κύκλο ζωής δεδομένων που περιέχει την άντληση, την επεξεργασία και τον διαμοιρασμό τους. Η Ευρωπαϊκή Ένωση λαμβάνοντας υπόψη τα πιθανά οφέλη τέτοιων πλατφορμών έχει, μέσω των σχετικών νομοθετικών πράξεων, προτείνει την ανάπτυξη Εθνικών Σημείων Πρόσβασης (ΕΣΠ) από κάθε Κράτος Μέλος. Το παρόν άρθρο έχει ως σκοπό να εντοπίσει τον ρόλο ενός ΕΣΠ μέσα στο οικοσύστημα των ΕΣΜ, να εξετάσει τις μεθοδολογίες σχεδιασμού ενός ΕΣΠ και μέσω της συγγραφής ενός εκτεταμένου σεναρίου χρήσης να αναδείξει τον τρόπο λειτουργίας ενός ΕΣΠ και να συσχετίσει πιθανά οφέλη με συγκεκριμένα στάδια της λειτουργίας του.

*Λέξεις Κλειδιά:* Ευφυή Συστήματα Μεταφορών (ΕΣΜ), Ανταλλαγή Δεδομένων, Εθνικά Σημεία Πρόσβασης, Εκτεταμένο Σενάριο Χρήσης, Ροή της Πληροφορίας, Δημιουργία Αξίας.

## *1. Introduction*

The transportation sector is continuously being reshaped in an effort to adapt itself to the context defined by the eminent proliferation of Information and Communication Technologies (ICT). In this respect, Intelligent Transportation Systems (ITS) may be considered as the main representative of this paradigm shift, recognizing that their vision is associated with the exploitation of ICT advancements for fully informing travelers, improving traffic safety, increasing transportation efficiency, reducing adverse impacts to the environment, while taking concurrently into account requirements related to seamless service provision, affordability, security, and privacy (Giannopoulos et al., 2012).

The extent to which the transportation sector has been adapted to this digital era, through the deployment of new ITS services, heavily relies on the availability of suitable and reliable data. To that end, data can be viewed as a 'new form of oil' for the sector's future trajectory (Catapult Transport Systems, 2015). However, several data-related limitations hinder the deployment of ITS services and the sector's overall potential for adaptation. These limitations range from the scarcity of available datasets and the lack of proper documentation to the sequestration of datasets in 'data silos' (Catapult Transport Systems, 2015; Jenkins, 2017).

Some of these limitations may be overcome with the deployment of various forms of data platforms in an effort to enable a seamless life-cycle of data harvesting, processing and display (Bacon et al., 2008). In addition to exposing data, a platform that includes data from many actors could further enhance the gathering of data as well as the benefits that arise from the processing of said datasets (Bacon et al., 2008). In an effort to capitalize on these benefits, various forms of data platforms have emerged ranging from simple open data portals all the way to complex and thorough digital marketplaces (Ayfantopoulou et al., 2019), each of which has its own uses and limitations.

In line with this concept, the European Union has proposed the deployment of National Access Point (NAPs) for the exchange and distribution of ITS data. This proposal was made within the context of the Delegated Regulations No. 885/2013, 886/2013, and 962/2015 that accompany the Directive 2010/40/EU on the framework for the deployment of Intelligent Transport Systems in the field of road transport and for interfaces with other modes of transport. According to these regulations, a NAP may be defined as a single digital interface at a national level, where data related to and derived from ITS are collected, properly formatted, enriched with the appropriate metadata, and made available to all interested parties. As of today, several Member States have started developing or are fully operating National Access Points. Figure 1 depicts a classification of the status of the Various NAPs as of 2018.

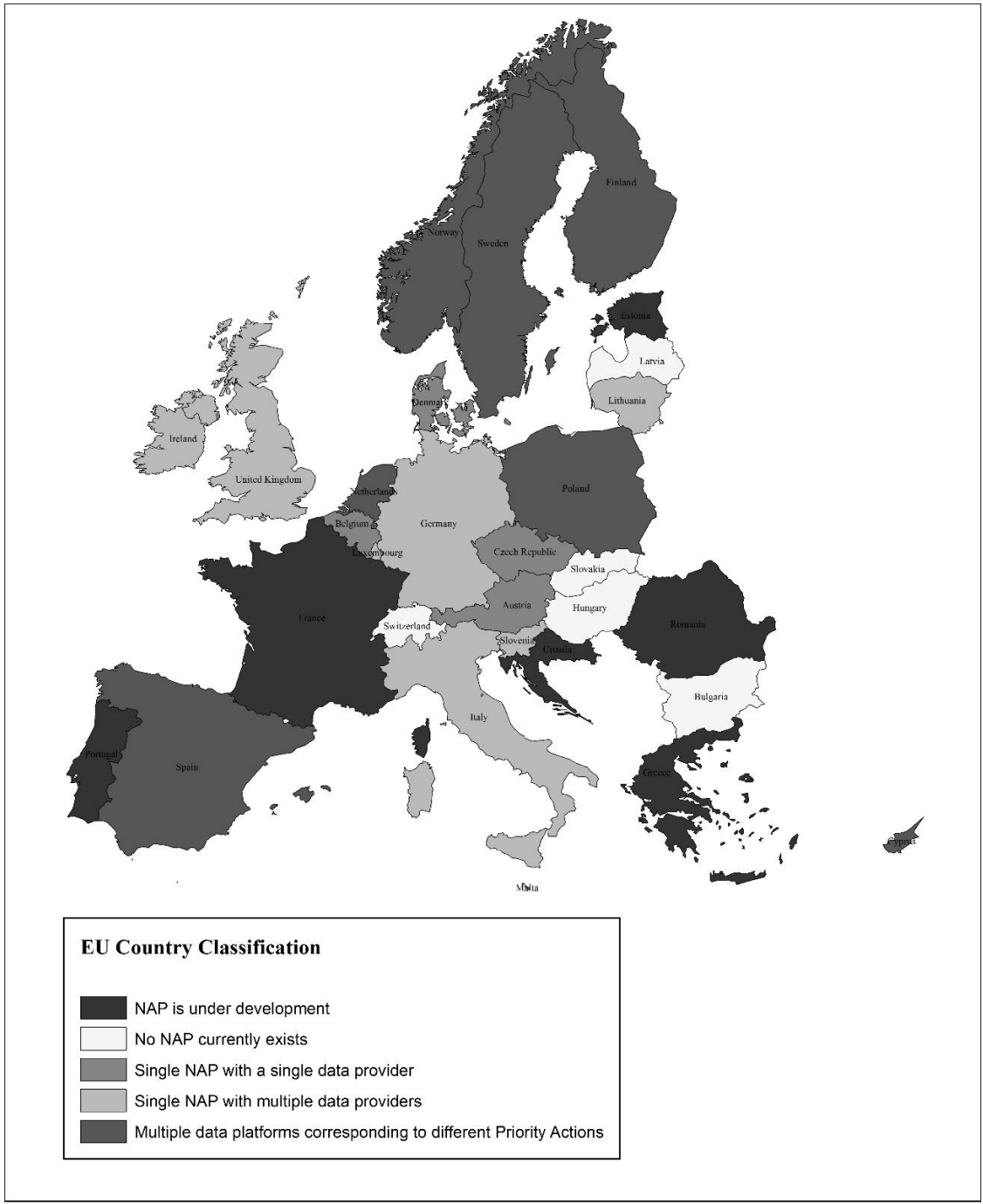

*__Figure 1:__ Classification of EU countries based on NAP status [Adapted from: Ayfantopoulou et al., 2019].*

The purpose of this paper is threefold. Initially, it aims to ascertain the importance of the various types of data for ITS and the transportation sector in general as well as the role of a NAP within the ITS ecosystem and the benefits it may provide to it. Secondly, this paper investigates alternative approaches to the nature and purpose of a data exchange platform and explores their compatibility with the purpose of a NAP taking into account its specific peculiarities. Thirdly, this paper demonstrates through an extensive use case the operational process of a NAP and juxtaposes specific steps of the process with identified benefits.

## *2. Research Methodology*

The first research objective is addressed through a review of the relevant literature. This is initially achieved by defining the concept of data within the context of this paper. Building on this analysis, this research objective hypothesizes the role of a NAP in dealing with specific types of data as well as bridging information gaps by providing meaningful connections between various types of data from various sources.

In order to address the second research objective, this paper investigates throughout the relevant literature the process being used to design data exchange platforms and possible variations of the final approach and design.

Finally, for the purposes of the third research objective, an extended use case is drafted and presented in two distinct variations. Through this analysis the operational process of a NAP is clearly presented and various identified benefits are associated with specific steps of the process in an effort to justify the need for the introduction of such a platform in the ITS ecosystem.

## *3. Data and Intelligent Transportation Systems*

Currently, the transport sector and especially road-based transportation faces a number of challenges, as seen in Figure 2. Within this figure, the increasing amount of data appears as a challenge, however, data may also be viewed as a means of empowering ITS and, therefore, as a part of the solution themselves. This may become more clearing when one takes into account the wide range of information that may be included in acquired data. Data within the context of ITS may inform responsible parties about the following (Ibrahim and Far, 2014):
- Traffic volume, speed, flow
- Vehicle occupancy
- Timestamp
- Road network type
- Scheduled and unscheduled events
- Accidents, road works and other hazards
- Archives of incidents
- Parking location, availability, and provided services
- Zoning (incl. areas of interest)
- Current and future weather conditions
- Glare (direction of sun)
- Road signage and sign locations
- Crowdsourced data (e.g. Tweets)

Apart from the nature of the provided information, the type of data through which it is provided is also of interest. The various types of data, as those are presented in Figure 3, may be classified according to either their spatial or temporal context (Chowdhury, Apon and Dey, 2017). It is of note that real-time data at a national level is of particular importance for ITS as is highlighted by the very proposed nature of NAPs.

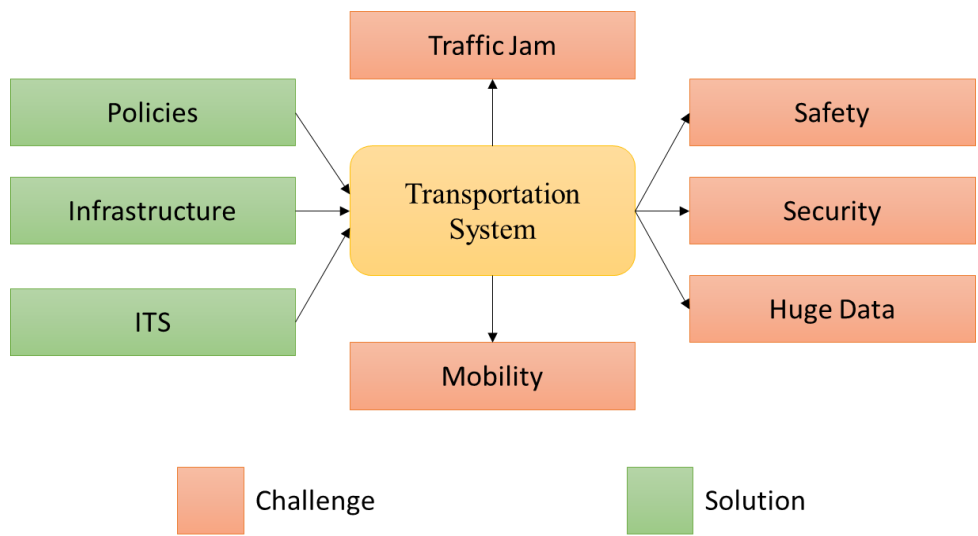

*Figure 2:* *Challenges and possible solutions within a transport system [Adapted from: Ibrahim and Far, 2014]*

Not only does the aforementioned variety of data produce, through the appropriate processes, useful information but they may also give rise to new function and services in ITS (Zhang et al., 2011). Some of the most prominent such applications of ITS that have been developed so far include, among other, services that allow for electronic payment of tolls as well as advanced: traffic management, vehicle and highway control, traveler and traffic information, public transportation (An, Lee and Shin, 2011). All of these services either require or produce data of a variety of types. Traditionally, the quality of such data is associated with their correctness, timeliness, and integrity (Staron and Scandariato, 2016). A more extensive list of the requirements of data would include the notions of believability, understandability, plausibility, sensitivity, timeliness, reputation, accessibility, appropriate amount, interpretability, integrity, trustworthiness, repudiation, and confidentiality (Staron and Scandariato, 2016).

The various types of data may be collected in a variety of ways and from a variety of sources. More specifically site-based data collection techniques utilize, for example, inductive magnetic loops, pneumatic road tubes, microwave radars, cameras and others (Prabha and Kabadi, 2016). Floating cellular data or floating car data (FCD) allow through triangulation, vehicle reidentification, or GPS-based methods (Prabha and Kabadi, 2016) the acquisition of varying types of information, such as direction, speed, travel times, congestion, etc. Finally, wide-area data collection, which may be achieved through satellite sensors, radio frequency identification techniques, or dedicated short range communications, may complement the advantages of FCD and other vehicle side technologies though the appropriate infrastructure (Prabha and Kabadi, 2016).

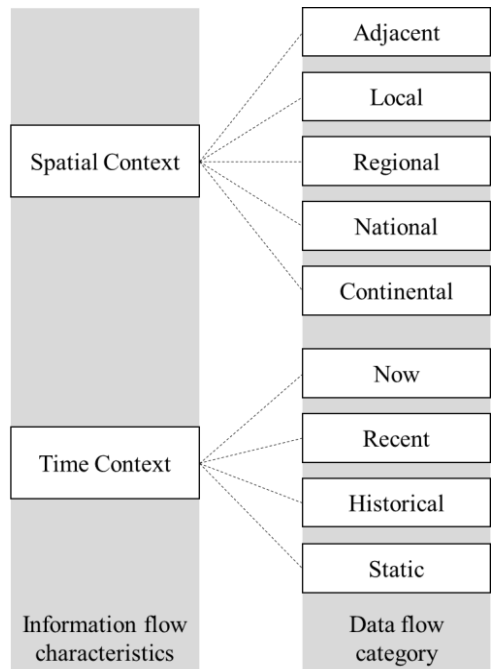

*Figure 3: Temporal and spatial distinction of ITS-related data [Adapted from: Chowdhury et al., 2017]*

However, it must be noted that the various types and sources of data have their own inherent strengths and weaknesses. For instance, FCD have the benefit of allowing the collection of information over a wide area and an initial low implanting cost; however, a low number of vehicles may represent the traffic stream and the measurement of day-to-day variability may be difficult (Toppen and Wunderlich, 2003). Similarly, while site-based data collection may allow for increased accuracy it provides limited coverage and has a higher implementation and maintenance costs (Prabha and Kabadi, 2016). The existence of varying benefits from and drawbacks in varying sources and types of data highlight an additional requirement of data, namely that of multi-sourcity.

Mutli-sourcity also becomes a requirement when one does not plan for a single app but rather for an ecosystem of app creation. This becomes apparent, as shown in Figure 4, in the fact that different types of apps or services require data of a varying nature.

Therefore, this multi-sourcity is essential and when combined with an increased rate of data gathering increase the value and potential of ITS. However, they also reveal the need for a common point of access through which the various actors may navigate this increasing complex data environment.

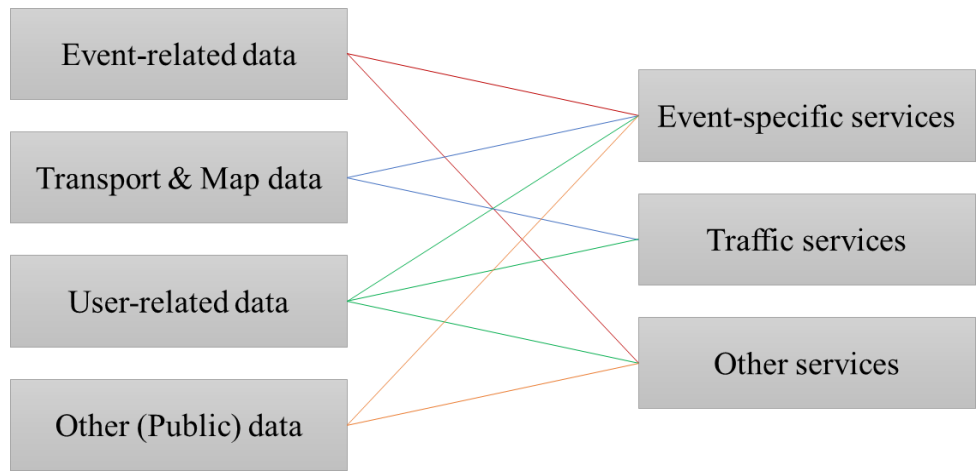

*Figure 4:* *Association of transport data with various services [Adapted from: Resch et al., 2008]*

## *4. The role of a NAP within the ITS ecosystem*

As already stated, the increasing requirements for and range of available data highlight the need for the establishment of a common point of access for new as well as previously undiscovered data. Such a point of access has been proposed and to some extent described in the Regulations (No. 885/2013, 886/2013, and 962/2015) that supplement the Directive 2010/40/EU (ITS Directive) as well as the relevant and associated literature. A broad definition is that such a focal node for ITS data that may be considered as a NAP has to have the characteristics of a digital point of access, where data are collected, properly formatted, associated with metadata, and made available for exchange and reuse (DG MOVE, 2018).

While the need for such platforms has been established, there exist and have been identified various barriers and challenges that may hinder their deployment and development. These barriers, as shown in Figure 5, highlight the need for a common approach in the design and development of NAPs across Europe.

Nonetheless, whichever approach is adopted by each country there are a series of benefits that have been associated with the introduction of a NAP within the ITS ecosystem. These benefits include the: a) improved comprehension of, b) improved processability of, c) improved discoverability of, d) improved (re)use of, e) increased trust towards, f) improved linkability between, g) greater interoperability of, and h) easier access to data (Ayfantopoulou et al., 2019). Two additional more generic benefits are included in this paper, namely those of an increase potential for innovation and greater operational efficiency of the road network.

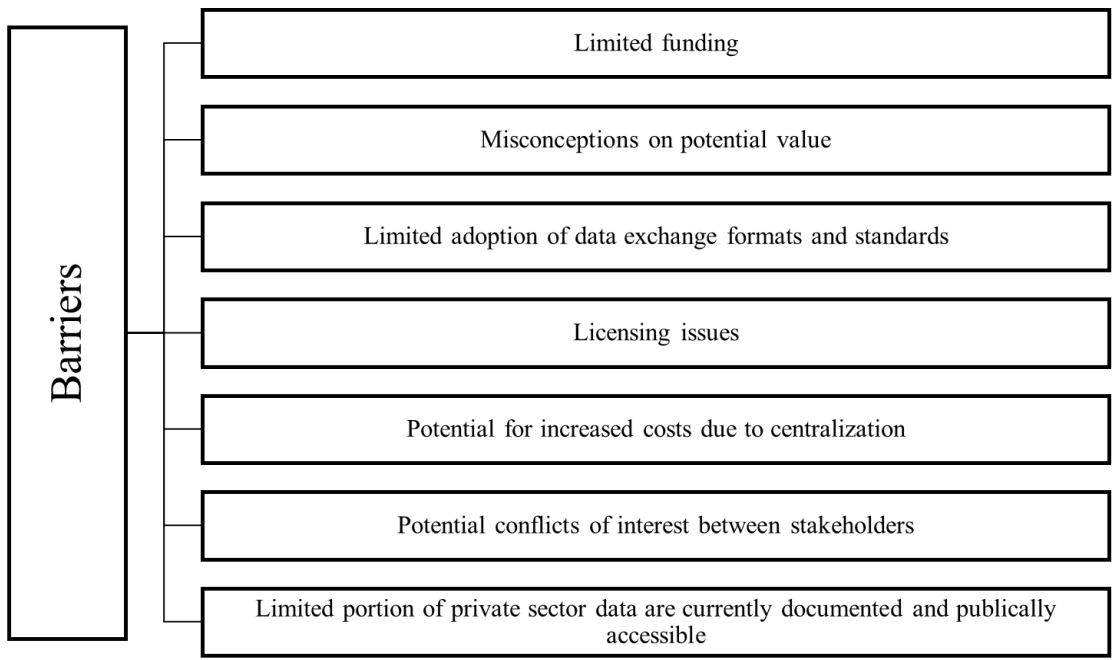

*Figure 5: Barriers to NAP deployment [Adapted from: Ayfantopoulou et al., 2019]*

The crucial role of a NAP in the ITS ecosystem is also expressed by the fact that the information flow in later generations of ITS is not as linear as in the past. Previously, competent road operators/ authorities used to monitor traffic in their own control centers by utilizing data deriving for instance from inductive loops or analogue cameras. This data were utilized in the proposal of appropriate measures for safety and fluency of traffic and their application via the roadside equipment (DATEX II, 2018). To the contrary, the recent years a lot of new actors has started to involve themselves in the provision of traffic information services (e.g. RDS-TMC). The provision of such services implies that these actors should be aware of traffic conditions, which means that they need data streams from the traffic control centers, in order to ensure the seamless and real-time provision of their services. In a further step and with the aim of widening the scope of the provided services, these actors have also started to combine data from various sources, which, as already stated in the previous section of this paper, implies the need for a common point of access to data as well as the need for data standardization. As shown in Figure 6, the solution of establishing a common point of access to the various sources is given by the ITS Directive through the creation of NAPs, while the solution for establishing a common language between data providers and service providers appears for now to be the DATEX II standard.

While DATEX was originally developed as means of standardizing the communication between TMCs and service providers, it is currently, in its second iteration, aiming to attract every possible actor in the dynamic traffic and traveler information sector under to adopt it (Rinne, 2014).

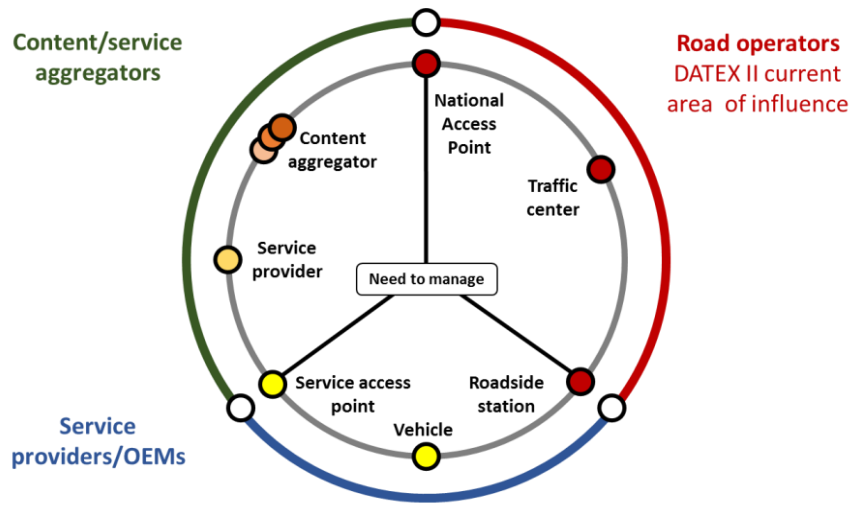

***Figure 6**: National Access Points and DATEX II as part of the ITS ecosystem [Adapted from: DATEX II, 2018]*

While the value of standardized formats such as DATEX has been recognized and their adoption is being promoted through the relevant regulations, their level of adoption varies based on factors such as technical readiness or nature of previously acquired and provided data. Figure 7 depicts the levels of DATEX II adoption throughout the European NAPs as of 2018.

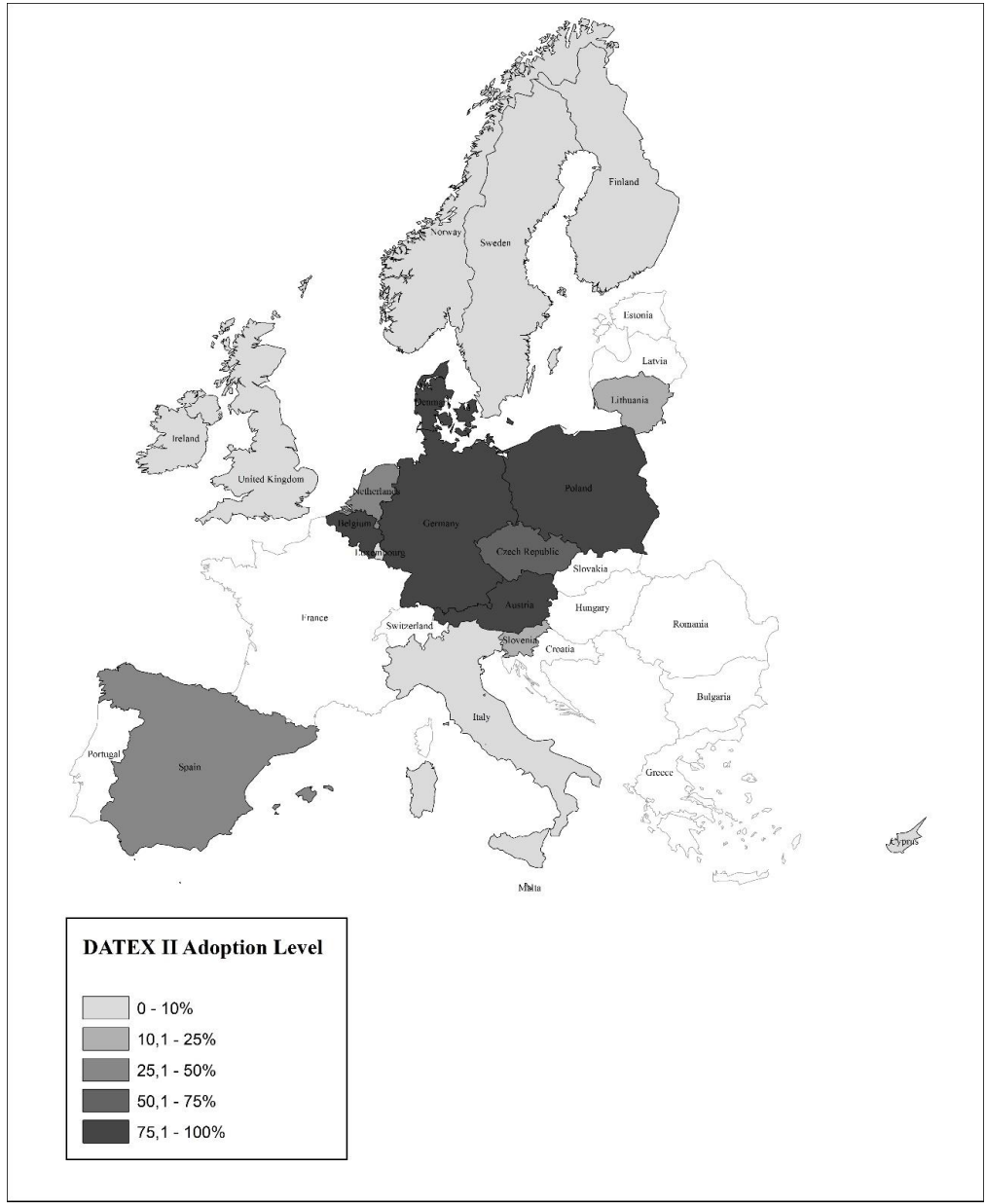

***Figure 7:*** *Level of DATEX II adoption per European NAP [Adapted from: Ayfantopoulou et al., 2019].*

## *5. Variations of data exchange platforms and design process*

As already stated, data exchange platforms may take a variety of forms. As Figure 8 depicts the bare minimum for such a platform is to serve as a raw data trader, working as an intermediary for raw data exchange. In a step above this, the platform would serve as a data normalizer by defining standard models, formats, and attributes, while verifying syntactically incoming data. Thirdly, such a platform may serve as a data aggregator by bundling data by logical classifications, such as region, transport mode or provider. Finally, and ideally, such a platform would serve as a quality assurer by carrying consistency and quality checks and rejecting invalid data. Any fees have to be proportional to the accountability for correctness of data (Deichmann, 2016).

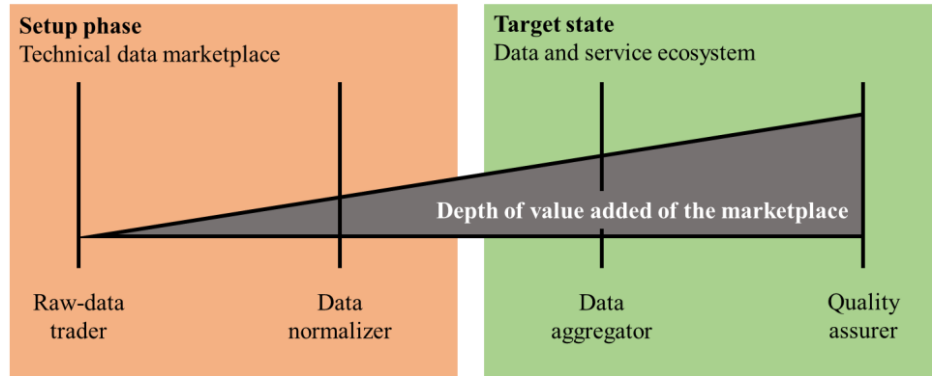

*Figure 8: Variations of data exchange platforms from least to most value creation [Adapted from: Deichmann, 2016]*

It should be noted that by legislating for the existence of National Assessment Bodies, the European Union defines NAPs as more closely resembling the concept of a quality assurer. At a bare minimum, a NAP has to serve as a data aggregator, in order to facilitate ease of access to a variety of data from a variety of sources.

Regardless of the type of data exchange platform in question a structured approach has to be adopted for its design and development. There are many such approaches, which include, amongst others, the waterfall, spiral, prototyping and V-model approaches (Figure 9).

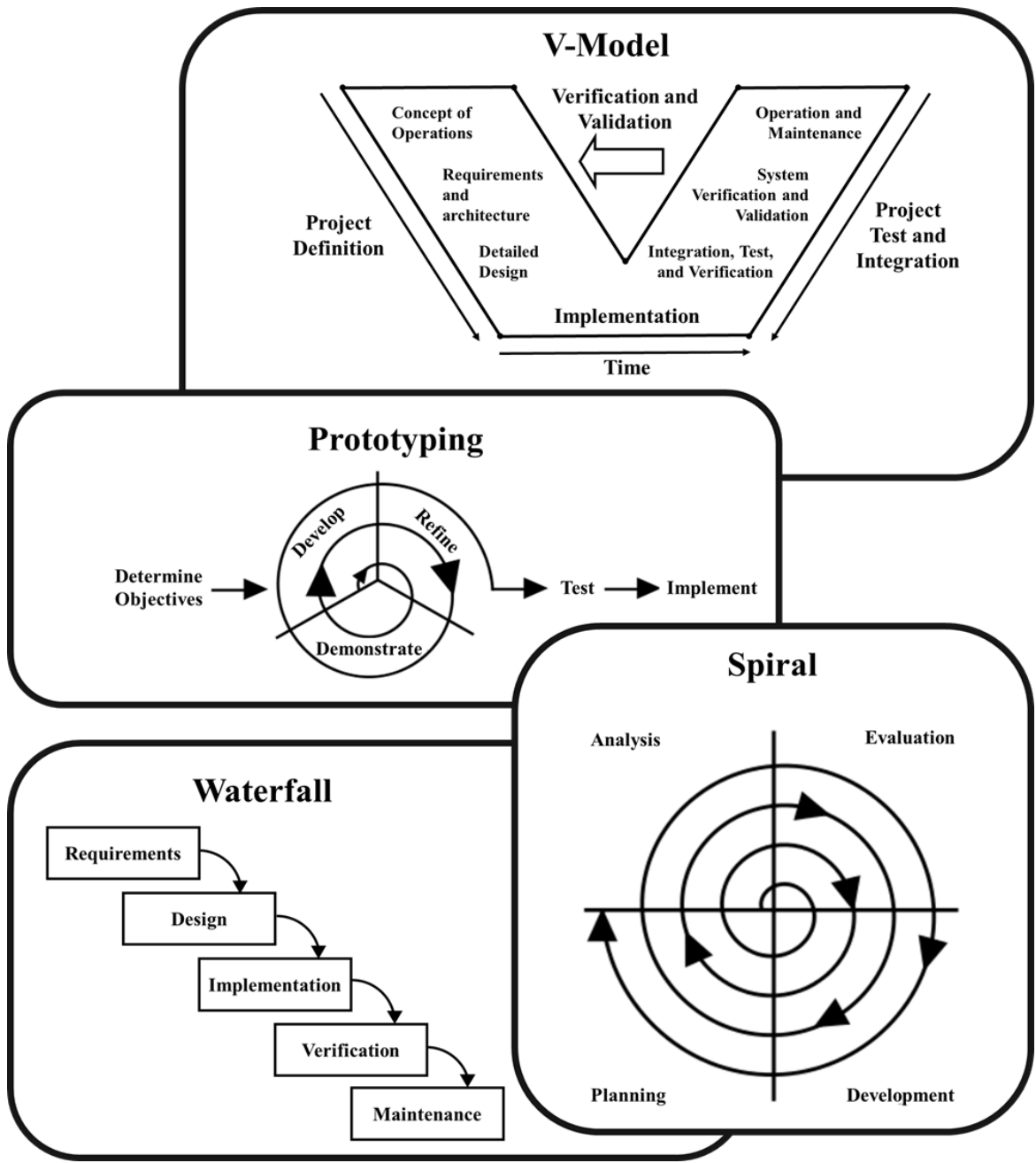

***Figure 9:*** *System development models [Adapted from: FHWA, 2005; Elgabry,2017]*

Whichever approach is chosen, any software and, therefore, data exchange platform design process must include the following four core elements (Elgabry, 2017):
- the specification or requirements engineering stage, where the main functionalities of the platform are defined;
- the design and implementation stage, where the platform is designed and programmed
- the verification and validation stage, where the platform is assessed against its specifications and requirements; and

- the maintenance and evolution stage, where changes to the platform may occur based on additional customer and market requirements.

By capturing both core and market requirements and identifying all of the platform's proposed features, one may more easily identify potential benefits, sources of added values and, therefore, business opportunities. The focus of the rest of this paper is to showcase such benefits and sources of added value in contrast with the operation process and the requirements of a NAP.

## *6. Demonstrating the operation of a NAP*

For the purposes of this section, a generic story is described and then presented in graphical form. This extended use case highlights core requirements and functionalities of a NAP at a software and business level.

"A truck driver is planning a long haul while carrying valuable items. They require assistance in planning their journey. To this end, they download a mobile application of assured quality, which acquires data available from multiple sources and which provides real-time information concerning road and weather conditions and safe and secure parking locations and availability."

This story may be simplified in the following two use cases based on whether a NAP exists or not:

- "User A (truck driver) acquires a service created by user B (ITS service developer), who uses data from multiple sources (e.g. Users C and D). Data are acquired by User B separately from each source and then validated and combined for the creation of the service."
- "User A (truck driver) acquires a service created by user B (ITS service developer), who uses data from multiple sources (e.g. Users C and D) and all of which are acquired via the NAP (User E) and validated by an Assessment Body (User F)."

The analysis of this story is presented in the form of Use Case Diagrams written in Universal Machine Language (Figure 10). This option was adopted as a supplement to textual descriptions as it may provide a quicker and clearer overview of the flow of events that occur throughout the story.

One of the first difference that one may observe by juxtaposing the two diagrams is the undue burden falling on the service provider, if no NAP exists. Characteristically, before creating their service User B has to locate Users C and D on their own, in order to inquire about the existence of and access to data. After coming to an arrangement User B has to assess the quality and suitability of the data and agree to their acquisition. After acquiring data from at least two sources User B has to harmonize potentially disparate data. This process may be required to take place repeatedly if data are deemed unavailable, unsuitable, or incompatible with those of other sources resulting in an exponential rise in complexity.

The same process, when a NAP exists, is simpler and more streamlined, even though additional actors are involved. Characteristically Users C and D access the NAP and create the appropriate Metadata by following the guidelines drafted by User E. They then publish their data on the platform. Data are in turn assessed by User F and harmonized by User E. User B now has to simply access the NAP, inquire for suitable data through the provided search mechanisms, and acquire them through the platform. Said data are guaranteed to be of adequate quality and compatible with one another, so under normal circumstances this process is only conducted once per required dataset.

The rest of the flow of events is identical for both versions of the use case, meaning that User B develops their service. Afterwards, User A acquires access to the service from User B and use the service to fulfill their needs.

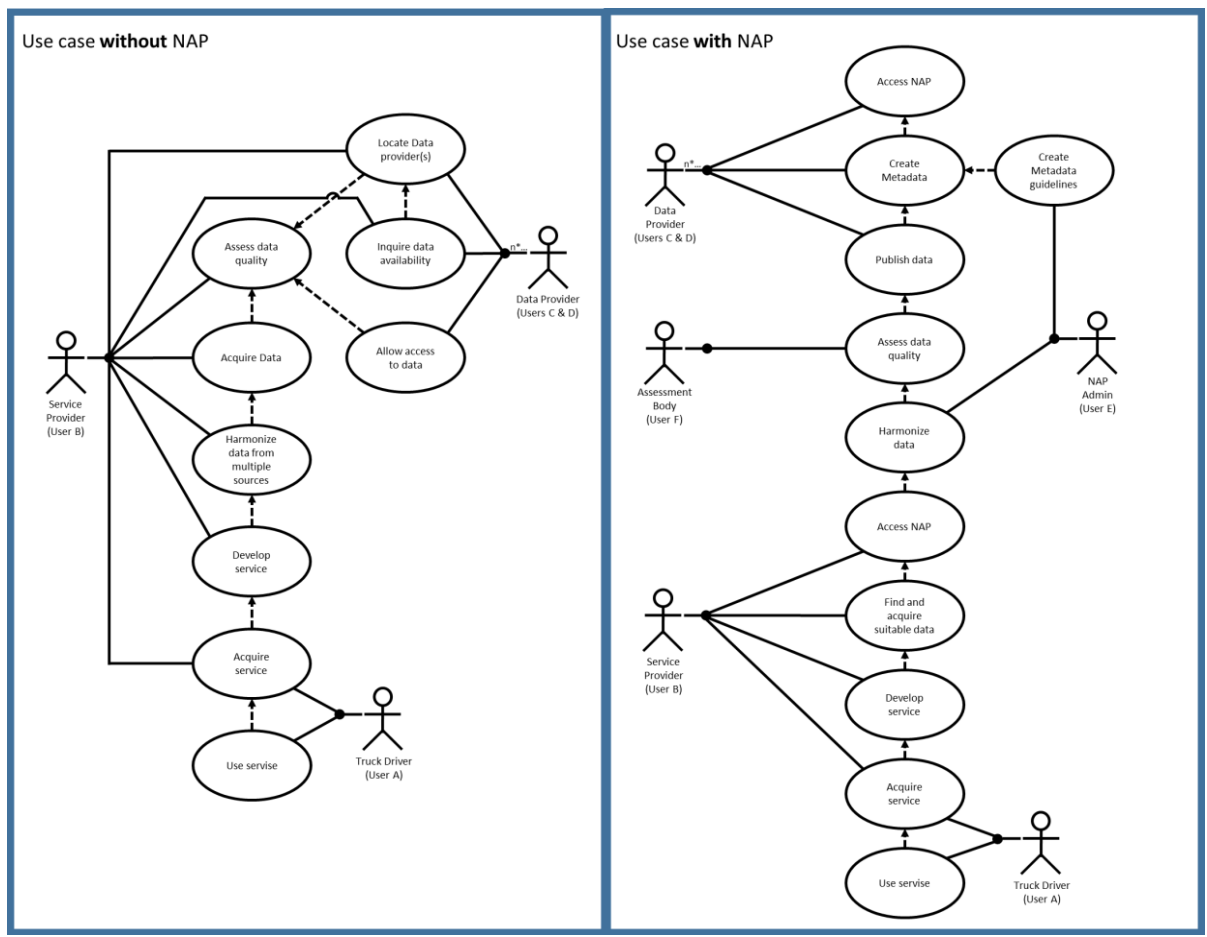

*Figure 10: Use case variations with and without the operation of a NAP*

## 7. Recognizing the benefits from the operation of a NAP

Upon a closer a closer inspection of the previous use case, one may notice besides the alleviation of excessive burdens on service providers, a series of additional benefits. Figure 11 describes benefits as those has been described in the relevant literature (Ayfantopoulou et al., 2019) and associates them with specific steps of the provided use case. This is expected to facilitate a better understanding of value creation and of the place of the NAP within the value chain.

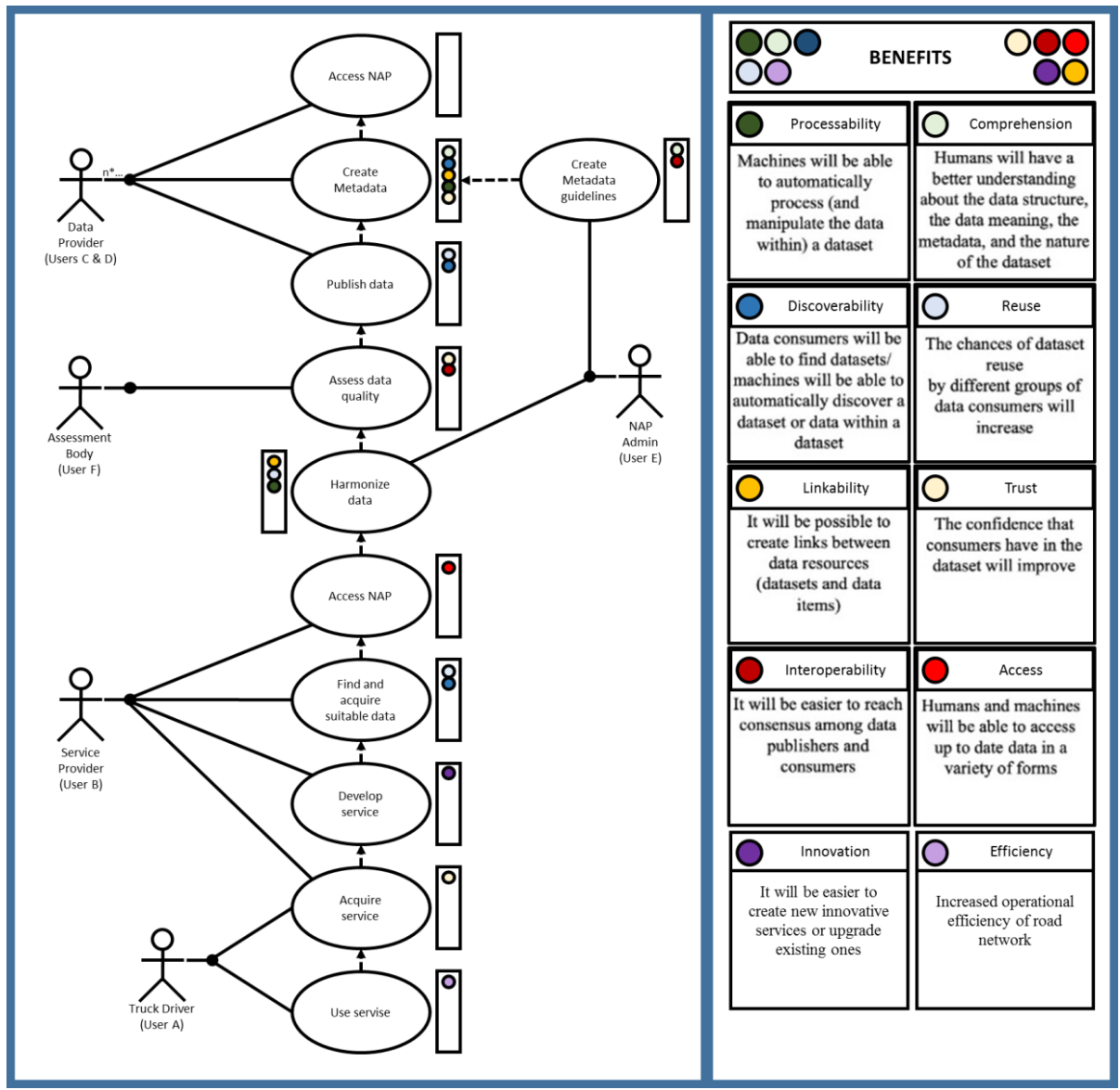

*Figure 11: Identification of NAP benefits and association with specific aspects of its operation*

The reasoning behind the association of benefits with each step of the use case is as follows:
- Access NAP (Data Provider): No benefits are associated with this step
- Create Metadata guidelines (NAP Admin): The first associated benefit is that of "comprehension" since it is assumed that the standardization of metadata forms the basis for a better understanding of the nature of data by all parties. The second associated benefit is that of "interoperability" since the creation of standardization guidelines by its very nature promotes the concept of a consensus between the various parties on the structure of data and metadata.

- Creation of metadata (Data Provider): This is perhaps the most beneficial step in the whole process, as it is associated with five distinct benefits. Firstly, metadata are expected to increase "comprehension" as they allow for a structured description of the content and nature of data. Furthermore, they facilitate "discoverability" since they may be used as tags and filters that allow for faster and more accurate inquires for data. Thirdly and in a similar manner, the same tags may be used as a bridge for linking different datasets since they help identify variables that may be merged. The fourth associated benefit is that of "processability" for without the existence of standardized metadata, it would be increasingly difficult for machines to identify and use data without the intervention of a human. Lastly, clear metadata are expected to promote "trust" towards the data they describe, since they promote transparency.
- Publish data (Data Provider): Two benefits are associated with this step, namely those of "reuse" and "discoverability". The association with the former is based on the fact that accumulation of multiple previously hidden datasets into a single point of access is by definition expected to increase their rate of use by new data consumers. Similarly, the association with the latter benefit is based on the fact that data are observable by the general public or dedicated machines only after they have been published.
- Assess data quality (Assessment Body): The importance of this step has been recognized by the European Union, which now requires of its Member States the formulation of national Assessment Bodies. It is, therefore, of no surprise that benefits are associated with this step. Firstly, and most importantly, this is the step that more acutely promotes the concept of "trust", since quality is now guaranteed by an official entity. Furthermore, conformity to the rules of set by a national Assessment Body is, even if forcibly, expected to promote consensus.
- Harmonize data (NAP Admin): Not all NAPs take on this responsibility. However, three benefits are associated with the NAP taking on the burden of harmonizing published data. Firstly, this process promotes the "linkability" of data from different providers, a fact that is of particular importance when one strives for cross-border continuity. Moreover, harmonized data are more easily read by machines and even by humans who now know precisely what to expect. In a similar manner, but to a lesser extent, this process also promotes "comprehension".
- Access NAP (Service Provider): Without this step it would be difficult to "access" data from multiple sources and in an appropriate manner.
- Find and acquire suitable data (Service Provider): This is the main step through which the "reuse" of data is accomplished, after taking advantage of the "discoverability" benefits intrinsic to process of looking for data through structured and user-friendly means.
- Develop service (Service Provider): This step represents the use of the NAP and its resources in an effort to achieve "innovation" meaning new or improved services.
- Acquire service (Service Provider; Road User): The only benefit associated with the transaction between end user and service provider is a potential increase in the "trust" towards data and their value, when new additional customers or users are attracted.
- Use service (Truck Driver): This step is associated with a perhaps wider benefit and one that encompasses the rest, namely the increased "efficiency" associated with the use of ITS-related services.

## *8. Conclusions*

This paper constitutes an initial attempt to explore the importance and the requirements of data for ITS. Furthermore, it explores the concept of a NAP and the possible variations it may take. It presents in a structured manner its operational process and further highlights its importance by juxtaposing the aforementioned process against one where on NAP exists. Furthermore, it associates the various identified benefits that have been proposed of it against specific steps of its operational process.

It is expected that in the following years most EU countries will have introduced national platforms of such nature. Furthermore, as standardization is expanded, with the increased levels of adoption of standards such as DATEX II, additional features such as efficient machine-to-machine communication are expected to become the norm.

Future publications are expected to expand with additional business-oriented use cases in an effort to capitalize on identified benefits in an efficient manner with the introduction of innovative business models.

## *9. References-Bibliography*